\documentclass[prb,10pt,floatfix,amsmath,twocolumn,superscriptaddress]{revtex4-1}
\usepackage[ansinew]{inputenc}
\usepackage{graphicx}
\usepackage{pst-all}
\usepackage{color}
\usepackage{dcolumn}
\usepackage{amsmath}
\usepackage{amsthm}
\usepackage{bm}
\usepackage{layout}
\usepackage{float}
\usepackage{txfonts}
\usepackage{amsfonts}
\usepackage{amssymb}%
\begin{document}
%\title{Design of hybrid Weyl/Dirac ferroelectric semimetal: a candidate of HgPbO$_3$}
%\title{Hybrid Weyl/Dirac ferroelectric semimetal}
%\title{Switchable Weyl/Dirac topological states coexisting with ferroelectriciy in the semimetal perovskite HgPdO$_3$}
\title{Weyl Ferroelectric Semimetal}

\author{Ronghan Li,$^{1,*}$ Yuanfeng Xu,$^{2,*}$ Jianggang He,$^{3,*}$ Sami Ullah,$^1$ Jiangxiu Li,$^1$ Jun-Ming Liu,$^4$ Dianzhong Li,$^1$
Cesare Franchini,$^{3,\ddagger}$ Hongming Weng,$^{2,5,\ddagger}$ and
Xing-Qiu Chen,$^{\ddagger,}$}

\affiliation{Shenyang National Laboratory for Materials Science,
Institute of Metal Research, Chinese Academy of Science, School of
Materials Science and Engineering, University of Science and
Technology of China, 110016 Shenyang, Liaoning, China, \\ $^2$
Beijing National Laboratory for Condensed Matter Physics, and
Institute of Physics, Chinese Academy of Sciences, Beijing 100190,
China, \\
$^3$ University of Vienna, Faculty of Physics and Center for
Computational Materials Science, Sensegass 8/8, A-1090 Vienna,
Austria,\\ $^4$ Laboratory of Solid State Microstructures and
Innovation Center of Advanced Microstructures, Nanjing University,
Nanjing 210093, China,\\
$^5$ Collaborative Innovation Center of Quantum Matter, Beijing,
China.}

\date{\today}

\begin{abstract}
{The recent discoveries of ferroelectric metal and Weyl semimetal
(WSM) have stimulated a natural question: whether these two exotic
states of matter can coexist in a single material or not. These two
discoveries ensure us that physically it is possible since both of
them share the same necessary condition, the broken inversion
symmetry.
%Here, we demonstrate that HgPbO$_3$ is such a unique material by using first-principles calculations.
%Noncentrosymmetric crystal might have ferroelectric
%instabilities and can also exhibit topological quantum phenomena such as
%\jh {chiral} Weyl nodes. However, ferroelectricity and topological Weyl
%semimetals \jh{(WSM)} have never been observed in a single common material,
%because ferroelectric behaviors generally develop in insulators
%whereas Weyl fermions are often formed in a semimetallic background.
Here, by using first-principles calculations, we demonstrate that
the experimentally synthesized nonmagnetic HgPbO$_3$ represents a
unique example of such hybrid ``\emph{Weyl ferroelectric
semimetal}''. Its centrosymmetric $R\bar{3}c$ phase will undergo a
ferroelectric phase transition to the ferroelectric $R3c$ structure.
Both phases are metallic and the ferroelectric phase owns a
spontaneous polarization of 33 $\mu$C/cm$^2$. Most importantly, it
also harbors six pairs of chiral Weyl nodes around the Fermi level
to be an oxide WSM. The structural symmetry broken phase transition
induces a topological phase transition. The coexistence of
ferroelectricity and Weyl nodes in HgPbO$_3$ is an ideal platform
for exploring multiphase interaction and mutual control. The Weyl
nodes can be tuned by external pulse electric field, which is
promising for potential applications of integrated topotronic and
ferroelectric devices.

%effects can be continuously tuned
%by gradually reducing the polar mode which restores the inversion
%center, thereby suppressing polarity, and leads to the disappearance
%of all Weyl nodes.
}
\end{abstract}

% \pacs{61.48.-c, 71.15.Mb,  73.22.-f}

\maketitle
\section{Introduction}
Noncentrosymmetric (NCS) lattice is an important asset for hosting
two seemly incompatible phenomena: ferroelectricity and topological
semimetals, such as WSM. Ferroelectricity arises from the formation
of electric dipoles due to the separated centers of positive and
negative charges under polar distortion, as shown in
Fig.~\ref{fig1}(a). Owing to the screening of free carriers in
metals, ferroelectricity and ferroelectric phase transition are
generally thought to occur only in insulators. However, in 1965 P.
W. Anderson and E. I. Blount suggested that, if a structural
transition occurs in a metal that is accompanied by the appearance
of a polar axis and the disappearance of inversion centre, a
``ferroelectric metal" can be expected~\cite{1}. Indeed,
ferroelectric metallic state and the ferroelectric phase transition
have been recently discovered in LiOsO$_3$~\cite{3} and several
other oxides~\cite{2,4,5}. On the other hand, NCS WSM with time
reversal symmetry have also been discovered~\cite{ts16,TSM,FS}. The
lack of inversion center breaks the Kramer degeneracy, leading to
separation of chiral Weyl nodes, which are formed by crossing of two
non-degenerate bands around Fermi level. Within this context, in
principle the ferroelectricity and WSMs can coexist in a single
material.

Although several nonmagnetic WSMs, such as TaAs
family\cite{taas1,taas2,taas3,taas4}, Ta$_3$S$_2$\cite{ta3s2},
TaIrTe$_4$ \cite{tairte4}, WTe$_2$\cite{wte2,wte3}, chemically doped
KMgBi \cite{Carrity2014,Narayan2015,Sante2016}, chalcopyrites like
CuTlSe$_2$\cite{cts}, HgTe class material~\cite{hgte}, and ZrTe
\cite{zrte} have been proposed and partially experimentally
confirmed, neither ferroelectricity nor ferroelectric phase
transition has been reported. According to the space group of these
crystals, the former five materials can have ferroelectricity
because the electric dipole is allowed to exist, whereas the latter
three cases are expected to have piezoelectricity due to the
nonexistence of polar axis. The ferroelectric phase transition is
accompanied with the inversion symmetry breaking and, hence, it is
also possible to induce topological phase transition to WSM since
the inversion symmetry and time reversal symmetry result in Kramer
degeneracy in bands and prevent the separation of Weyl nodes with
opposite chirality.

%\xq{Recently, a class of ferroelectric materials with polar
%hexagonal semiconductors, hyperferroelectrics~\cite{Carrity2014},
%was theoretically suggested to be coupled with strong Rashba effect
%\cite{Carrity2014,Narayan2015,Sante2016}. Furthermore,
%hyperferroelectrics were unveiled to be a topological insulator and,
%even, show a phase transition from a topological insulator to a Weyl
%semimetal and eventually to a normal insulator by chemical doping
%and alloying~\cite{Sante2016}}.While conceptually possible, the major obstacle is the
%the practical realization of such a material and design rules are
%necessary.

%Weyl semimetals represent a novel quantum state exhibiting
%fundamentally new properties such as exotic nontrivial Fermi arcs
%and unusual transport properties, that are functional for
%applications in topological superconductivity, quantum computing,
%and spintronics \cite{ts16}. Weyl semimetallic states were recently
%discovered in the NCS-structured TaAs
%family~\cite{taas1,taas2,taas3,taas4}, but they can be also realized
%in Dirac semimetals~\cite{TDS,OC1,OC2}, in which spin-doublet
%degenerate Dirac cones occur in the Brillouin zone (BZ), just by
%breaking time reversal or spacial inversion symmetry, like in
%Na$_3$Bi\cite{na3bi-1,na3bi-2,na3bi-3,na3bi-4} and
%Cd$_3$As$_2$\cite{cdas1,cdas2,cdas3,cdas4}, respectively.

To form ferroelectric metal it is necessary to have a finite
occupation at the Fermi level and an electronic structure
insensitive to polar ionic displacement in a polar
lattice~\cite{1,2,3,4,5}, thus ensuring that orbital-like electronic
states at the Fermi level are decoupled from the polar displacement.
On the other hand, the design principles for WSMs include a fully
filled band at the border of a semiconducting regime and a strong
spin-orbit coupling (SOC) which could cause large Rashba-like spin
splitting in a NCS lattice. By combining these design principles we
achieve the following criteria for the realization of coexisting
Weyl nodes and ferroelectric metal: (\emph{i}) a polar and flexible
lattice, (\emph{ii}) fully filled electron orbitals, and
(\emph{iii}) strong SOC. Clearly, oxides represent the most
promising playground owing to their chemical and structural
flexibility and to the abundance of data available in
literatures~\cite{Halasyamani1998, cal2}. To fulfill the
fully-filled orbital condition we adopt the simple chemical concept
of electronegativity as used in our previous work~\cite{Lirh2015}.
Finally, the strength of SOC can be estimated from the atomic number
of the atomic constituents. We find that these conditions are
simultaneously satisfied in the rhombohedral oxide HgPbO$_3$. Based
on first-principles calculation we demonstrate that HgPbO$_3$ is a
WSM with ferroelectric phase transition, that we name ``\emph{Weyl
ferroelectric semimetal}''.

\section{Computational Methods}
1) {\bf Structural optimizations and electronic band structures} The
structural optimization and electronic properties were performed
within the framework of the density function theory (DFT)
\cite{1964-DFT,1965-DFT} by employing the Vienna \emph{ab initio}
simulation package(VASP)\cite{1993-Kresse,1996-Kresse} based on the
projector augmented wave (PAW) method\cite{1994-Blochl}. We have
used the generalized gradient approximation of
Perdew-Burke-Ernzerhof (PBE) \cite{1996-Perdew} The adopted PAW-PBE
pseudopotentials of Hg, Pb and O treat 5$d^{10}$6$s^1$,
5$d^{10}$6$s^2$6$p^2$ and 2$s^2$2$p^3$ electrons as valence states.
The energy cutoff was set at 500~eV and appropriate Monkhorst-Pack k
meshes (13$\times$13$\times$13) were chosen. A very accurate
criteria were used for the structural relaxation (forces $<$ 0.0001
eV/\AA\,). In order to double check the electronic band structures
with the presence of Dirac and Weyl nodes, we have also employed the
hybrid density functional theory developed by Heyd, Scuseria and
Ernzerhof (HSE)\cite{HSE1,HSE2} and the GW method \cite{GW}. For
HSE, we have adopted the standard set-up for $\mu$ and $\alpha$:
$\mu$ = 0.20 \AA\,$^{-1}$ controls the range separation between the
short-range and long-range parts of the Coulomb kernel and whereas
$\alpha$ = 0.25 is the mixing parameter determining the fraction of
exact Hartree-Fock exchange incorporated.

2) {\bf Phonon calculations} To inspect the dynamical stability of
the ground state structure, we carried out phonon calculations. The
phonon properties were computed using finite-displacements and
density functional perturbation theory (DFPT)~\cite{2001-Baroni}, in
supercells containing 2$\times$2$\times$2 primitive cells. The
phonon spectra was derived by interfacing VASP with the code
Phonopy\cite{2008-Togo}.

3) {\bf Surface electronic band structures} The surface state
calculations have been performed using a Green's function based
tight-binding (TB) approach~\cite{GF}. The TB model Hamiltonian was
constructed by means of maximally-localized Wannier functions
(MLWFs)\cite{Wannier1,Wannier2} obtained by the the Wannier90
code\cite{2008-Mostofi} and constructed from \emph{s-, p-, d}
atomic-like orbitals. The TB parameters were obtained from the MLWFs
overlap matrix.

4) {\bf Electric polarization calculations} The spontaneous electric
polarization is estimated by
\begin{math} P_{\alpha}=
\frac{e}{\Omega}\sum\limits_{k,\beta}Z_{k,\alpha\beta}^*u_{k,\beta}\end{math},
using the normal charges ($Z^*$ , Hg: +2; Pb: +4; O: -2) and atom
displacements ($u$) of the polar structure ($R{3c}$) with respect to
the reference phase $R{\bar{3}c}$, where $\Omega$ and $e$ are the
volume of the unit cell and the elementary electron charge,
respectively.

\section{Results and discussions}

\begin{figure}[!h]
\centering
\includegraphics[width=0.48\textwidth]{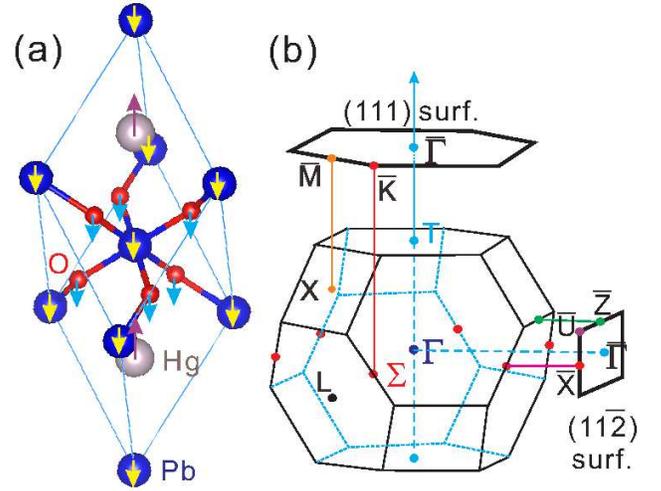}
\caption{{\textbf{Crystal structure and Brillouin zone (BZ) of
HgPbO$_3$}. Panel (a) denotes the unit cell of centrosymmetric
$R\overline{3}c$ rhombohedral lattice.
%: the optimized lattice constants are $a$= 5.8112 \AA\, and $c$ = 14.4639 \AA\; Hg is located at the 2$a$ site (0.25, 0.25, 0.25), Pb at the 2$b$ site (0, 0, 0) and O at the 6$e$ site (0.1268, 0.76, 0.37318).
The arrows at each atom denote the atomic displacement in the
softest phonon mode $\Gamma_{2}^{-}$ with imaginary frequency of
1.28$i$ THz at $\Gamma$ in Fig.~\ref{fig2}. Such polar distortion
leads to its ferroelectric phase of $R3c$ lattice.
%the optimized lattice constants are $a$ = 5.8014 \AA\, and $c$ = 14.5865\AA\; Hg is located at the 2a site (0.2358, 0.2358, 0.2358), Pb at the 2a site (0, 0, 0) and O at the 6b site (0.1296, 0.7829, 0.3755).
Panel (b) shows the BZ and high symmetrical k-points of the
rhombohedral lattice, as well as the projected (111) and
(11$\bar{2}$) surfaces BZ.
%which are equivalent to the (0001) and (11$\bar{2}$0) surfaces of the hexagonal unit cell of HgPbO$_3$. Noted that the DFT derived lattice constants are in nice agreement with the experimentally measured values~\cite{exp1} ($a$ = 5.7515$\pm$0.0005 \AA\, and $c$ =14.534$\pm$0.001 \AA\,).
}} \label{fig1}
\end{figure}

\begin{figure}[hbt]
\centering
\includegraphics[width=0.48\textwidth]{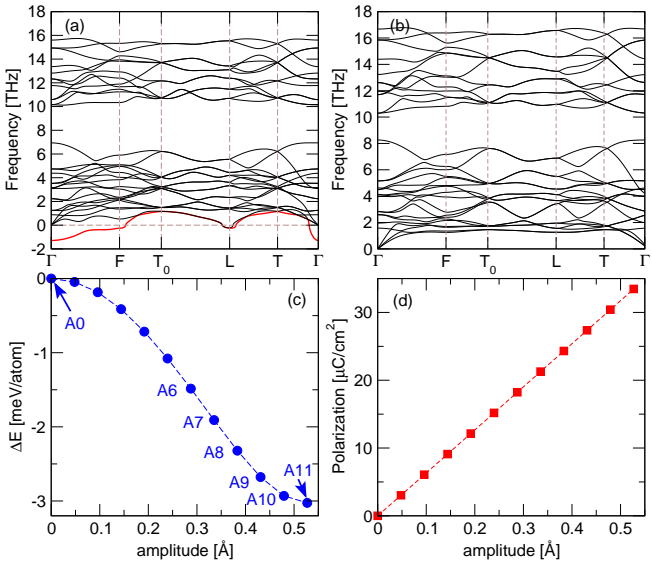}
\caption{{\textbf{Phonons and ferroelectric distortion of
HgPbO$_3$.} Calculated phonon dispersions for the nonpolar
$R\bar{3}c$ (a) and polar $R3c$ (b) structures. (c) The energy
difference $\Delta{E}$ between the polar and nonpolar phases and (d)
the electric polarization $P$ as a function of the amplitude (A0,
..., A11) of the frozen polar mode $\Gamma_2^{-}$.
%The high symmetry points are $\Gamma$ (0, 0, 0), $F$ (-$\pi$, $\pi$, 0), $T_0$ (-$\pi$, $\pi$, $\pi$), $L$ ($\pi$, 0, 0) and $T$ ($\pi$,$\pi$, $\pi$), respectively.
}} \label{fig2}
\end{figure}

\begin{figure}[hbt]
\centering \vspace{0.1cm}
\includegraphics[width=0.48\textwidth]{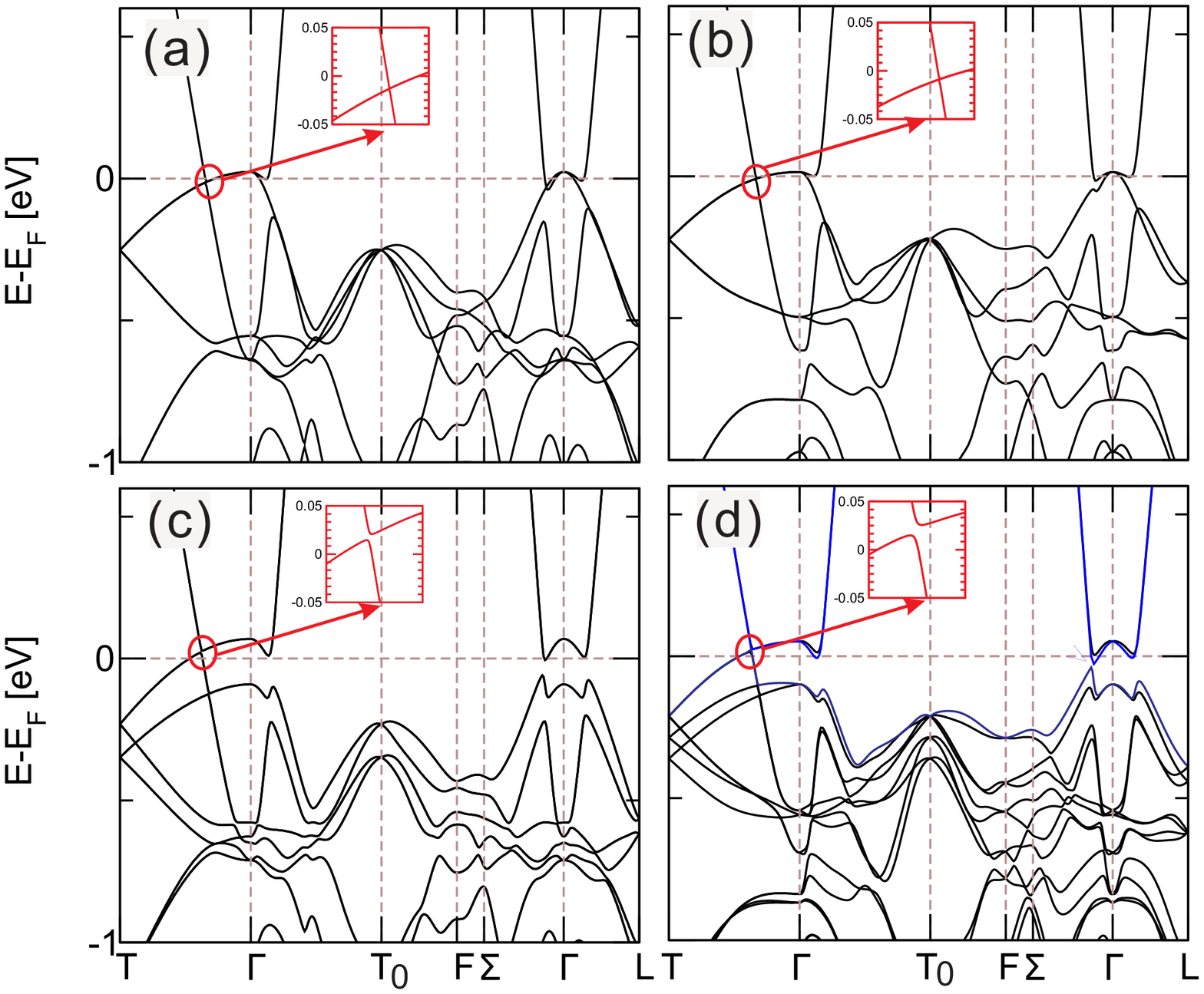}
\caption{{\textbf{Calculated electronic band structures of
HgPbO$_3$.} Panels (a) and (b) are those calculated without SOC for
the nonpolar $R\bar{3}c$ and polar $R3c$ phases, respectively.
Panels (c) and (d) are the corresponding ones with SOC included. The
insets stress the crossing (without SOC) and anti-crossing (with
SOC) of the bands mainly composed of 6$s$ orbitals of Hg and Pb
atoms and 2$p$ orbitals of oxygen atoms.
%It can be seen that, regardless of SOC, the six-fold spin
%degeneracy along the T-$\Gamma$ direction is always present in
%panels (a) and (b). The inclusion of SOC breaks the six-fold spin
%degeneracy into a tiny gap in panels (c) and (d) for both $R3c$ and
%$R\bar{3}c$ phases. Note that electronic structures obtained using
%the more accurate hybrid functionals (HSE) and GW methods are almost
%identical to the DFT one (\jh{supplemental Fig. S3 and Fig.S4}).
}}~\label{fig3}
\end{figure}

\begin{figure}[hbt]
\centering \vspace{0.1cm}
\includegraphics[width=0.48\textwidth]{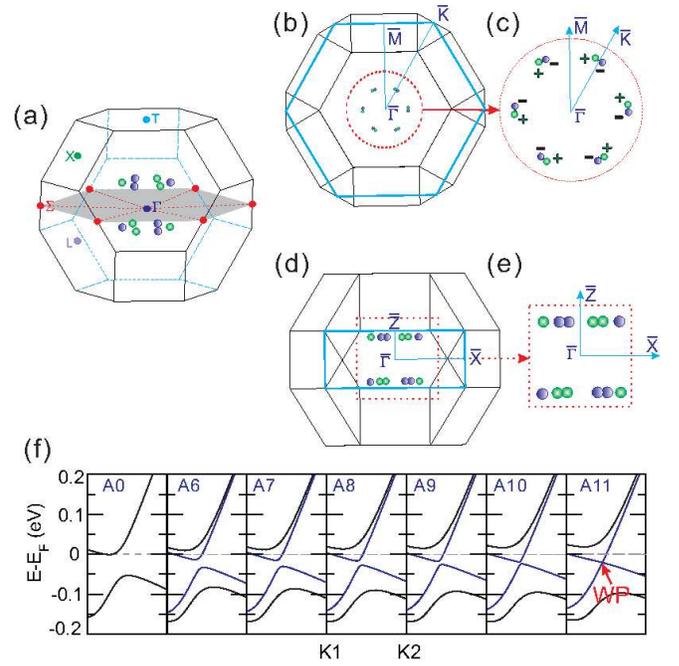}
\caption{{{\bf Weyl nodes and their distribution in the BZ of the
most stable ferroelectric $R3c$ HgPbO$_3$.} One of the Weyl node is
at (0.15273470, 0.06175918, -0.01762040) and all the others can be
obtained by imposing crystal and time-reversal symmetries (see the
supplementary Table S1).
%marked by WP in panel (a) along the K1 (0.12773469, 0.06175918, 0.0073796)
%to WP (0.15273470, 0.06175918, -0.01762040) to K2 (0.17773469, 0.06175918, -0.0426204) directions.
Panel (a): three-dimensional visualization of the Weyl nodes in the
BZ, (b) two-dimensional projection along the $<$111$>$ direction and
the panel (c) is the enlarged plot of the projected Weyl nodes, and
(d) two-dimensional projection along the $<$11$\bar{2}$$>$ direction
and the panel (e) is the enlarged plot of the projected Weyl nodes.
The green and blue points denote the opposite chirality of the
corresponding nodes. Panel (f) shows the changes in the band
structure at one Weyl node as the function of the amplitude of polar
distortion (A0, ..., A11 as labeled in Fig. \ref{fig2}c) along the
path parallel with the $\bar{\Gamma}$-$\bar{K}$. }}~\label{fig4}
\end{figure}

HgPbO$_3$ was synthesized in 1973 in the temperature range 600-1000
$^\circ$C under 30 to 65 kbars pressure~\cite{exp1}. Unlike the
isoelectronic CdPbO$_3$ which was found to be a polar
semiconductor~\cite{cal1,cal2}, HgPbO$_3$ was reported to be black
and weakly conducting ($\rho$ $\approx$ 10$^{-3}$
ohm/cm)~\cite{exp1}. Experimental ambiguity on the  crystal
structure remains, specifically on whether HgPbO$_3$ is polar
($R$3$c$) or nonpolar ($R\overline{3}c$), see Fig.
\ref{fig1}(a)~\cite{cal2}. To clarify this issue we have conducted a
direct comparison between the two phases and found that the $R3c$
phase is more stable than $R\overline{3}c$ by 3 meV/atom. Further
support for this conclusion is provided by the calculated phonon
dispersions. As shown in Fig.~\ref{fig2}, the nonpolar
$R\overline{3}$c phase has imaginary frequencies whereas the polar
$R3c$ phase does not have and is dynamically stable.

The most unstable phonon mode of the $R\overline{3}c$ phase is the
infrared active mode of $A_{2u}$ at $\Gamma$ ($\Gamma_2^-$), which
involves the cooperative displacements of the negatively charged
oxygen ions opposite to the positively charged Hg and Pb cations
along the $c$-axis of the hexagonal lattice, or equivalently the
$<$111$>$ direction of rhombohedral lattice, as shown in
Fig.~\ref{fig1}. By freezing the $\Gamma_2^{-}$ mode at different
amplitude, the centrosymmetric $R\overline{3}c$ lattice is shown to
be transformed into the polar $R3c$ lattice with energy gain of 3
meV/atom (Fig. \ref{fig2}c) at the ground state. All of these
distorted crystals are metallic in their band structure, reflecting
well the experimentally measured weak conductivity \cite{exp1}.
%A well localized charge redistribution associated with the
%formation of local dipole moment \jh{(supplementary Fig.S1)} clearly indicates a ferroelectric
%phase transition, with the polar axis $c$.
By using the normal charges of cations and anion (Hg: +2; Pb: +4; O:
-2), the estimated spontaneous polarization along the polar axis $c$
is as high as 33~$\mu$C/cm$^2$. The rather small energy change (3
meV/atom) during this process implies an accessible ferroelectric
phase transition with decreasing temperature. This is consistent
with the analysis of Anderson and Blount~\cite{1} for the
identification of a ferroelectric metal. Similar to the prototype
ferroelectric metal LiOsO$_3$, the main polar distortion arises from
opposite displacements of the $A$-site cation (Hg) (0.37 \AA) and O
atoms (0.27 \AA) along the $c$-axis. The $B$-site (Pb) cation has
the smallest displacement (0.1 \AA). Considering the large polar
distortion (0.53~\AA in amplitude of $\Gamma_2^-$ mode), small
energy difference $\Delta{E}$ between the polar and nonpolar phase,
and weak screening in semimetal of low carrier density, the polar
ground state is expected to be tunable by applying suitable pulse
electric field opposite to the 'spontaneous polarization'.

The inspection of the electronic structures shows that in HgPdO$_3$
the ferroelectric instability is coupled with striking topological
phase transition from a normal semimetal to a WSM with Weyl nodes,
as
evidenced in Figs.~\ref{fig3} and \ref{fig4}. %The driving force
%guiding the onset of these unusual behaviors is the SOC arising from
%the heavy Hg and Pb atoms.
By neglecting the SOC, the band structure of the $R\bar{3}c$ and
$R3c$ phases are very similar, characterized by a band crossing
around the Fermi level along the $\Gamma$-T direction. One of the
band is a nondegenerate band mainly composed of 6$s$ orbitals from
Hg and Pb atoms, while the others are double degenerate bands mostly
of oxygen atoms' 2$p$ orbitals. Such $s$-$p$ band inversion
(supplementary Fig. S1) leads to a triply degenerate nodal point
along $\Gamma$-T, similar to that in TaN and others~\cite{TaN, TPTM,
zrte}. However, such triply degenerated nodal point becomes gaped
once SOC is further included. In the polar $R3c$ case, the gap is
about 7 meV and in the centrosymmetric $R\overline{3}c$ case the gap
is smaller, about 5 meV. It needs to be emphasized that electronic
structure obtained using the more accurate hybrid functionals (HSE)
and GW methods is almost identical to the DFT one (supplemental Fig.
S2 and Fig. S3). Another most important difference between these two
phases is that the Kramer degeneracy is broken in ferroelectric one.
The Rashba like splitting can be clearly seen when comparing the
bands around $\Gamma$ along either $\Gamma$-$\Sigma$ or
$\Gamma$-$L$. This might lead to appearance of separated chiral Weyl
nodes in NCS polar $R{3}c$ phase. And indeed, we have identified
that there are totally six pairs of Weyl nodes in the first BZ as
shown in Fig.~\ref{fig4}. Each of them is related with the others by
$C_3$ rotational symmetry around the $c$ axis, the mirror symmetries
with mirror planes passing through $\Gamma$-T and $\Gamma$-L, and
the time-reversal symmetry. The energy of the Weyl node is around 22
meV below the Fermi level. The exact position and the chirality of
these Weyl nodes are specified in supplementary Table S1. The Weyl
cone shape can be seen from the energy dispersions along several
different directions as shown in supplementary Fig.S4. Again, we
monitor the evolution of band structure around one Weyl node as the
amplitude of the polar distortion of $\Gamma_2^{-}$ soft mode
changes. As clearly shown in Fig.~\ref{fig4}, the topological phase
transition happens during the ferroelectric phase transition. Thus,
we have identified an unusual and tunable coupling between
topological Weyl fermion and the ferroelectric properties.

\begin{figure}[hbt]
\centering
\includegraphics[width=0.49\textwidth]{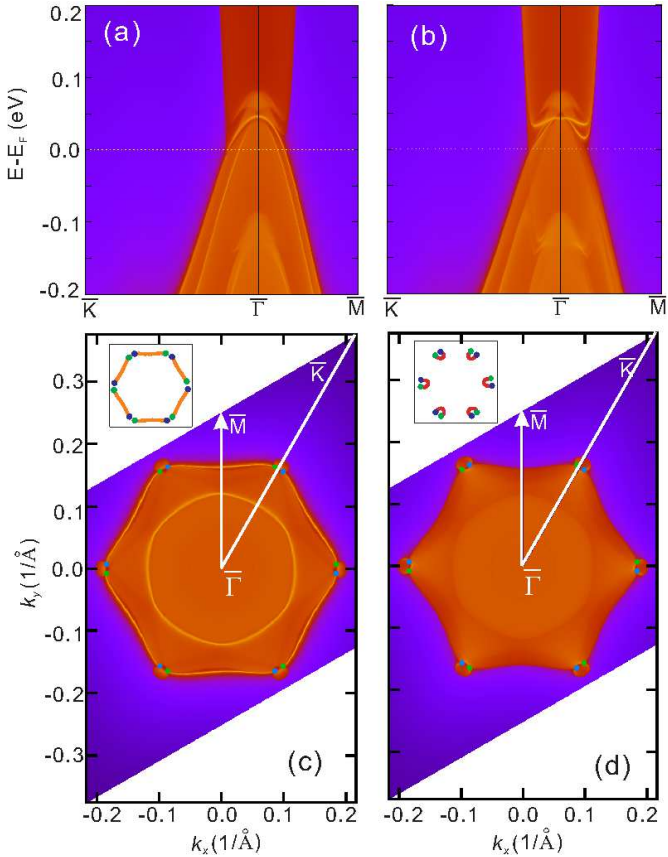}
\caption{{\textbf{Calculated (111) surface state of the polar $R3c$
HgPbO$_3$}. Panels (a) and (b) are the band structures of the top
and bottom surfaces, respectively. Panels (c) and (d) are the Fermi
surfaces at -22 meV below the Fermi level in panels (a) and (b),
respectively. Insets in (c) and (d) show the Fermi arc pattern. Blue
and green points are the surface projections of the Weyl nodes of
the opposite chirality. }} \label{fig5}
\end{figure}

\begin{figure}[hbt]
\centering \vspace{0.1cm}
\includegraphics[width=0.49\textwidth]{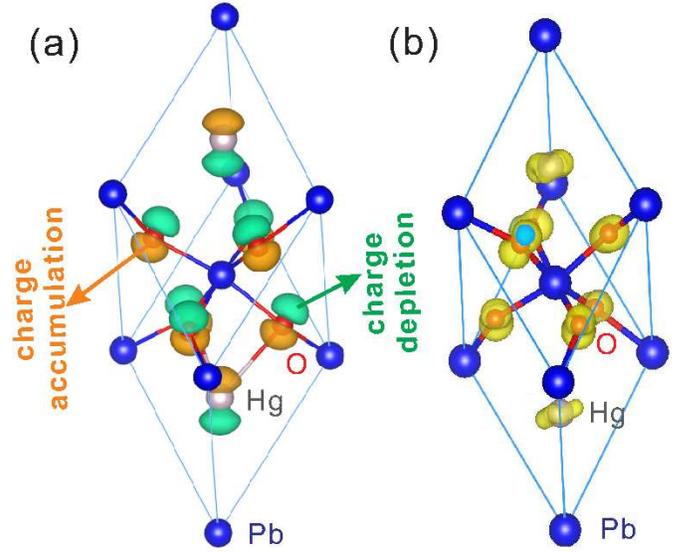}
\caption{{\textbf{Localized charges associated with local electric
dipoles and Weyl carriers of the polar $R3c$ HgPbO$_3$} Panel (a)
Charge distribution difference,
$\Delta{\rho}=\rho_{R3c}-\rho_{R\bar{3}c}$, between that of the
polar $R3c$ ($\rho_{R3c}$) and nonpolar $R\bar{3}c$
($\rho_{R\bar{3}c}$) structure of HgPbO$_3$. (b) Distribution of the
carriers in the electron-pockets containing Weyl nodes. The orange
and green colors denote the charge accumulation and depletion,
respectively. It is clearly to see that the $\Delta{\rho}$ on Hg and
O are accumulated and depleted in the opposite direction to form
dipole moment because of the polar displacement.}}~\label{fig6}
\end{figure}

WSM often exhibits nontrivial surface states~\cite{ts16,TSM}, such
as Fermi arcs. To inspect them for the ferroelectric HgPbO$_3$, we
calcualte the surface electronic band structures within the
tight-binding model based Green's function method. The Wannier
functions are generated from the the bulk Hg 6$s$, Hg 5$d$, Pb 6$s$
and O 2$p$ orbitals as the basis set of tight-binding model.
Fig.~\ref{fig5} shows the (111) surface band structures and their
corresponding Fermi surfaces at the energy of Weyl nodes for the top
and bottom surfaces. As illustrated in Fig.~\ref{fig5}a and
\ref{fig5}b, the projected Weyl nodes are hidden by the other
overlapped bulk states. But there are still some surface states with
quite high surface local density of states can be identified.
Fig.~\ref{fig5}c and \ref{fig5}d are the plotting of Fermi surface
with chemical potential at Weyl nodes. The Fermi arcs connecting the
projection of the Weyl nodes with opposite chirality are well
resolved for both top and bottom surfaces. We have also checked the
surface electronic bands on the (11$\bar{2}$) top and bottom
surfaces, perpendicular to the (111) surface (see supplementary
Fig.~S5(a and b)). However, all these projected Weyl nodes on the
(11$\bar{2}$) surface still overlaps the bulk bands as illustrated
in supplementary Fig.~S5(c and d).
%Clearly, by reducing the amplitude of the polar mode $\Gamma_2^{-}$
%the intensity of the Fermi arc states would progressively disappear,
%reflecting the gradual attenuation of the Weyl nodes, and the Dirac
%cones would be clearly visible in surface band structure projected
%in the (111) plane.

Finally, to reveal the underlying mechanism of the coexistence
between WSM and ferroelectricity in HgPbO$_3$, we find that the
charges at the Weyl nodes mainly contribute to well localized charge
redistribution associated with the formation of local dipole
moments, as shown in Fig.~\ref{fig6}. It is clear that
ferroelectric-like phases are not expected to form in metals, mainly
because the screening potential imposed by free electrons in metals
prevents the electric dipoles from feeling each other and, hence,
stops them from aligning in order~\cite{Lezair2016}. However,
compared to standard metals, WSM appear to be a more suitable
platform to host ferroelectric metal, since the charge carriers in
Weyl nodes predominantly move along the linearly dispersive bands
forming the nodes. Due to this confinement, Weyl carriers are not
expected to be easily redistributed in all the lattice space but
localized on the orbitals composing the linear bands. This is
different from the nearly free electrons with parabolic band
dispersion in normal metals, where the electrons can travel over all
the lattice space to be extended. Moreover, the strength of
electrostatic screening in WSM is much weaker than the normal metal
due to the much lower carrier density. Therefore, if a NCS lattice
of WSM is flexible enough to allow the cooperative atomic
displacement with the positively and negatively charged ions to form
electric dipoles in a proper ferroelectric pattern, this most
probably will form a Weyl ferroelectric semimetals.

\section{conclusions}
In conclusion, we have shown that HgPbO$_3$ is a semimetal with
ferroelectric phase transition and its dynamically stable $R3c$
phase has dipole moment. Notably, such ferroelectric metal possesses
six pairs of Weyl nodes, which represents the first example of an
oxide Weyl ferroelectric semimetal. The Weyl nodes can be
selectively tuned by applying an external pulse electric field,
implying potential applications as quantum switch in processing
devices. The coexistence and the coupling between WSM phase and the
ferroelectric phase transition, assisted by strong relativistic
effects in a NCS background, has been achieved by judicious material
design based on well-defined building criteria. It represents a
unique feature never observed hitherto in known Weyl
semimetals~\cite{Wybeyond} and ferroelectric metals~\cite{5}. This
opens the way to novel quantum devices based on topological
multifunctional materials.

\bigskip
\noindent {\bf Acknowledgments} Work in IMR, China, was supported by
the ``Hundred Talents Project'' of the Chinese Academy of Sciences
and the Major Research Plan (Grant No. 91226204), the Key Research
Program of CAS (Grant No. KGZD-EW-T06), the National Natural Science
Foundation of China (Grant Nos. 51671193, 51474202 and 51431006) and
the computational resources from the national supercomputer center
(TH-2A). Y. X. and H.M.W in IOP, China, are supported by the
National 973 program of China (Grant No. 2013CB921700), the
``Strategic Priority Research Program (B)" of the Chinese Academy of
Sciences (Grant No. XDB07020100), the MOST project under the
contract number 2016YFA0300604 and the National Natural Science
Foundation of China (Grant Nos. 11274359, 11422428 and 11674369).
C.F. acknowledges support by the Austrian Science Fund (FWF) project
INDOX (I1490-N19) and ViCoM (Grant No. F41). Calculations in Vienna
were performed in the Vienna Scientific Cluster.

\bigskip
\noindent $^*$These authors contributed equally to this work.\\
$^\ddagger$ Corresponding author:{\\xingqiu.chen@imr.ac.cn (X.-Q. C.),\\
cesare.franchini@univie.ac.at (C. F.), \\ hmweng@iphy.ac.cn (H. M.
W.)}

\clearpage
\begin{table*}
\footnotesize \caption {Supplementary Table S1: Positions and
chirality of six pairs of Weyl nodes in the first Brillouin zone.
The position ($k_x$, $k_y$, $k_z$) are in units of reciprocal
lattices. All of them are related by crystal and time-reversal
symmetries.}
\begin{ruledtabular}
\begin{tabular}{crrrc}
Weyl point & $k_x$ & $k_y$ & $k_z$ & Chirality \\
\hline
W1 &  0.15273470 &  0.06175918 & -0.01762040  & -\\
W2 &  0.01762040 & -0.06175918 & -0.15273470  & +\\
W3 & -0.01762040 &  0.15273469 &  0.06175918  & -\\
W4 & -0.15273470 &  0.01762040 & -0.06175918  & +\\
W5 &  0.06175918 & -0.01762040 &  0.15273469  & -\\
W6 & -0.06175918 & -0.15273649 &  0.01762040  & +\\
W7 & -0.15273469 & -0.06175918 &  0.01762040  & -\\
W8 & -0.01762040 &  0.06175918 &  0.15273469  & +\\
W9 &  0.01762040 & -0.15273649 & -0.06175180  & -\\
W10&  0.15273469 & -0.01762040 &  0.06175918  & +\\
W11& -0.06175918 &  0.01762040 & -0.15273469  & -\\
W12&  0.06175918 &  0.15273469 & -0.01762040  & +\\
\end{tabular}
\end{ruledtabular}
\label{tab:Tab1}
\end{table*}

\begin{figure*}[hbt]
\centering \vspace{0.1cm}
\includegraphics[width=0.78\textwidth]{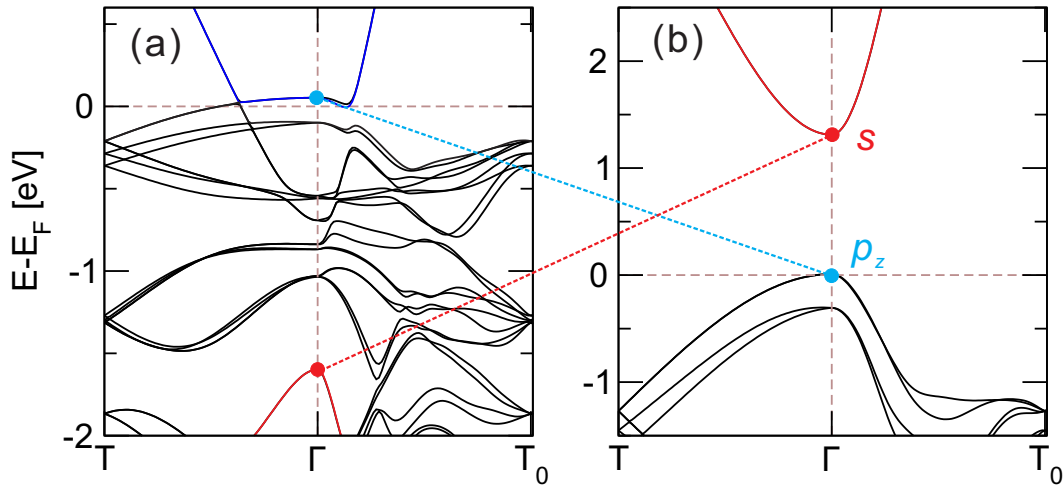}
\caption{Supplementary Fig.S1. {\textbf{$s \rightarrow p$ band
inversion.} To elucidate band inversion mechanism, we have analyzed
the band sequencies of the $R3c$-type HgPbO$_3$ at the equilibrium
state (a) in comparison with the artificial state by reducing the
volume to 50\% of the equilibrium volume. It can be seen that, at
the equilibrium state the band structure shows a semimetal, whereas
at the reduced volume it becomes an insulator. In the meanwhile, the
band inversion occurs between the \emph{s}-like orbitals from Hg and
Pb atoms and the $p$-like orbitals from oxygen
atoms.}}~\label{figs1}
\end{figure*}

\begin{figure*}[hbt]
\centering \vspace{0.1cm}
\includegraphics[width=0.78\textwidth]{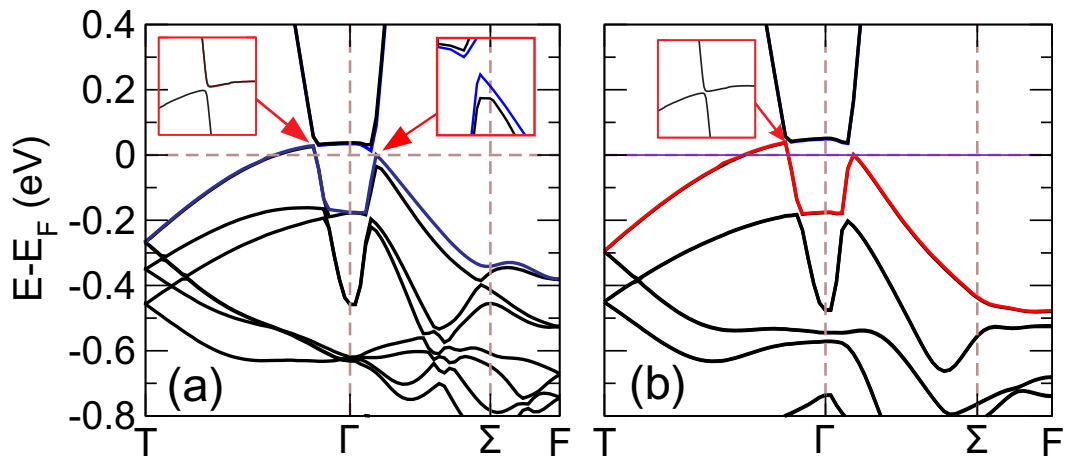}
\caption{Supplementary Fig.S2. {\textbf{HSE derived electronic band
structures of HgPbO$_3$.} Due to the well-known false-positive
problem of standard DFT calculations, HSE method with SOC has been
performed to check whether the topologically non-trivial states in
HgPbO$_3$ still exists when a more accurate exchange-correlation
functional is introduced. The derived band structure of the NCS
$R3c$ and centrosymmetric $R\bar{3}c$ structures in (a) and (b),
respectively, are very similarity to the DFT
calcualtions.}}~\label{figs2}
\end{figure*}

\begin{figure*}[hbt]
\centering \vspace{0.1cm}
\includegraphics[width=0.78\textwidth]{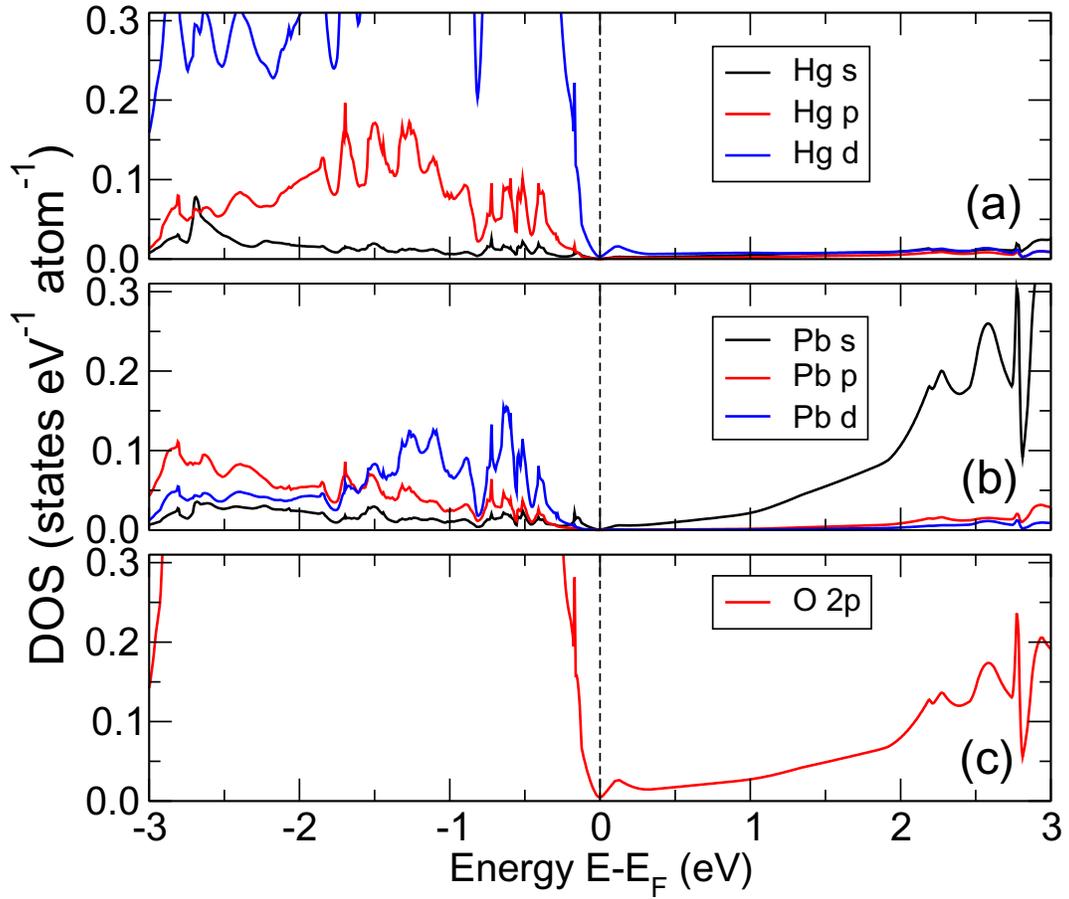}
\caption{Supplementary Fig.S3. {\textbf{GW derived electronic
densities of states (DOSs) of $R3c$ HgPbO$_3$.} Because HgPbO$_3$ is
semimetal, the GW method is not a most proper method to derive its
electronic band structures. However, according to our calculations,
the GW method still revealed that the GW DOSs show a clear sign of
the semimetal feature, as evidenced by a highly low density at the
Fermi level.}}~\label{figs3}
\end{figure*}

\begin{figure*}[hbt]
\centering
\includegraphics[width=0.8\textwidth]{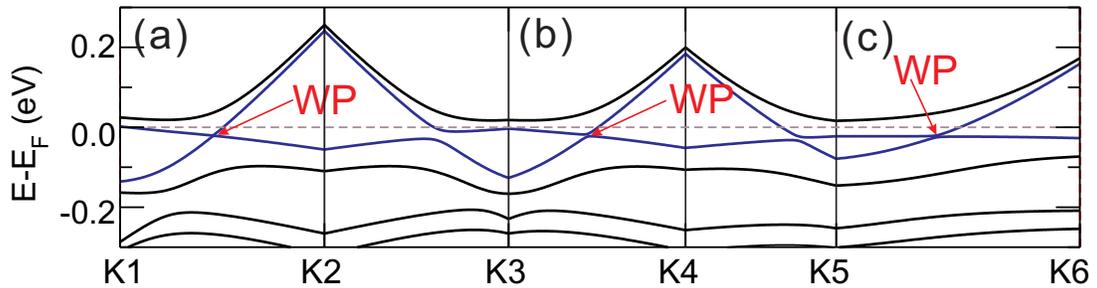}
\caption{Supplementary Fig.S4. {\textbf{Calculated band structure
along three directions passing through one Weyl node in the polar
$R3c$ phase of HgPbO$_3$}. Panels (a), (b) and (c) are those with
k-path parallel to $\bar{\Gamma}$-$\bar{K}$,
$\bar{\Gamma}$-$\bar{M}$ and $\Gamma$-T, respectively.}}
\label{figs4}
\end{figure*}

\begin{figure*}[hbt]
\centering
\includegraphics[width=0.6\textwidth]{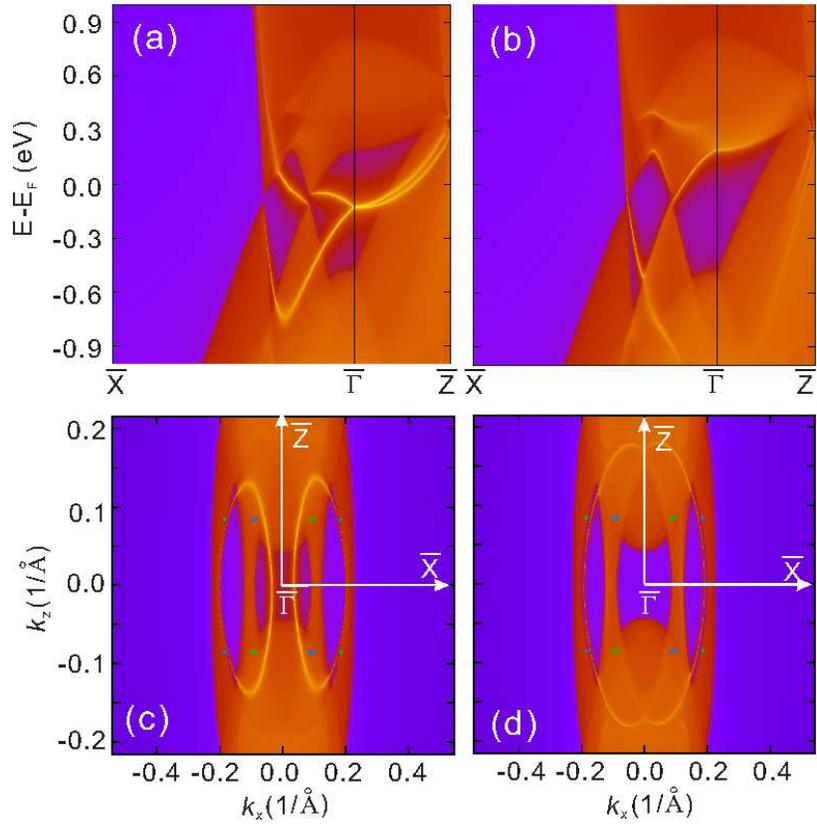}
\caption{Supplementary Fig.S5. {\textbf{Calculated
(11$\overline{2}$) surface electron structures for the polar $R3c$
phase of HgPbO$_3$}. Panels (a) and (b) are the calculated band
structures of the top and bottom surfaces, respectively. Panels (c)
and (d) are the Fermi surfaces at -22 meV below the Fermi level in
panels (a) and (b), respectively. Blue and green points are the
surface projections of the Weyl nodes of the opposite chirality.}}
\label{figs5}
\end{figure*}

\end{document}